\documentclass[12pt]{article}
\usepackage{graphicx,cite,epsfig}
\newcommand{\ba}{\begin{array}}
\newcommand{\ea}{\end{array}}
\newcommand{\bd}{\begin{displaymath}}
\newcommand{\ed}{\end{displaymath}}
\newcommand{\be}{\begin{equation}}
\newcommand{\ee}{\end{equation}}
\def\bt{\begin{table}}
\def\et{\end{table}}
\def\bi{\begin{itemize}}
\def\ei{\end{itemize}}
\def\bea{\begin{eqnarray}}
\def\eea{\end{eqnarray}}

\def\N0{\widetilde{\chi}^0}

\def\g{\gamma}

\def\m{\mu}

\def\s{\sigma}

\def\t {\times }

\def\h{\phi^{\pm\pm}}
\def\f{\phi^{--}}     
\catcode`@=11 
\def \gsim{\mathrel{\mathpalette\@versim>}}
\def \lsim{\mathrel{\mathpalette\@versim<}}
\def \@versim#1#2{\lower0.4ex\vbox{\baselineskip\z@skip\lineskip\z@skip
     \lineskiplimit\z@\ialign{$\m@th#1\hfil##\hfil$%
     \crcr#2\crcr\sim\crcr}}}
\catcode`@=12 
\begin{document}
\setcounter{page}{0}
\thispagestyle{empty}
\begin{flushright}
HRI-P-05-08-001 \\
\end{flushright}
\begin{center}
{\Large\sc Associated single photons as signals for a doubly 
charged scalar at linear $e^- e^-$ colliders} \\
\vspace*{0.2in}
{\large\sl Biswarup Mukhopadhyaya {\rm and} Santosh Kumar Rai} \\
\vspace*{0.3in}
Harish-Chandra Research Institute, \\
Chhatnag Road, Jhusi, Allahabad, \\ 
India 211 019. \\ 
Electronic address: {\sf biswarup@mri.ernet.in, skrai@mri.ernet.in} \\
\vspace*{1.3in}
{\Large\bf ABSTRACT}
\end{center}
\begin{quotation} \noindent \sl
\noindent
Doubly charged scalars, predicted in many models having
exotic Higgs representations, can in general have
lepton-number violating (LFV) couplings. The basis of most searches for this
charged scalar has been to look for its direct production and its 
subsequent decay to like-sign final state leptons. In this work we show
that by using an associated monoenergetic final state
photon seen at a future linear $e^- e^-$ collider, we can have a clear
and distinct signature for a doubly-charged resonance and also determine
its mass rather precisely. We also estimate the strength of the $\Delta
L=2$ coupling which can be probed in this way at $\sqrt{s}=1$ TeV, as
a function of the recoil mass of the doubly-charged scalar.
\end{quotation} \rm\normalsize
\vfill
\newpage

The absence of the 
Higgs boson from the cupboard containing {\it particle trophies} 
discovered by experimentalists, still leaves a scope of speculation as 
to what would eventually be the possible structure for 
$SU(2)\times U(1)$-breaking. Thus scenarios with extended Higgs sectors,
ranging from ones with two or more doublets to those with other
representations of SU(2), are often considered. 
Doubly charged scalars arise in a number of such scenarios \cite{THM,LRSM}.
The most common models to accommodate such scalars 
are those with triplet Higgs. Triplets  can be made part of the electroweak 
symmetry breaking sector in purely phenomenological studies
\cite{Georgi,BRD,dchpheno}. On the other hand,
they may be indispensable in some special theories such as the simplest
versions of  Little Higgs models \cite{LH}, where the Higgs is 
envisioned as a pseudo-goldstone boson, and the triplets are required to 
cancel the quadratic divergence to the Higgs mass.

An added feature often associated with doubly-charged Higgs is the possibility
of lepton-number violation. This basically consists in $\Delta L =2$ 
couplings with leptons of the form

\bea
 \mathcal{L}_Y = i h_{ij}\Psi^T_{iL}C\tau_2\Phi\Psi_{jL} + h.c.
\eea
where $i,j=e,\mu,\tau$ are generation indices, the $\Psi$'s are the 
two-component left-handed lepton fields, and $\Phi$ is the triplet with 
$Y=2$ weak hypercharge and is given by the $2\t2$ matrix of the scalar 
fields:
$$\left(\begin{array}{cc}
\phi^+/\sqrt{2}&\phi^{++} \\
\phi^0         &-\phi^+/\sqrt{2} \end{array}\right)
$$

This leads to mass terms for
neutrinos \cite{THM,srikanth} once the neutral component 
$\phi^0$ acquires a vacuum expectation value (vev):

\begin{equation}
\mathcal{M}^\nu_{ij}  \sim h_{ij} v'
\end{equation}

\noindent
$v'$ being the triplet vev.
 Since constraints on the $\rho$-parameter \cite{PDG} puts strong limits
on the the triplet vev \cite{LEP} in general,  this immediately translates 
to limits on the L-violation Yukawa couplings from the expected ranges of
neutrino masses \cite{Barger}. Such limits usually 
constrain the collider signals for 
doubly-charged scalars sought through $\Delta L = 2$ interactions. 
Of course, there are models where the limits from the $\rho$-parameter
can be avoided \cite{Gunion} by postulating real as well 
as complex triplets at the same 
time, and assuming a custodial symmetry relating their vev \cite{Georgi}. 
Although
such a custodial symmetry has been found to be preserved in higher-order
corrections from the scalar potential \cite{Chanowitz}, its stability 
against corrections 
via gauge coupling is not obvious. Thus it is safe to abide by
the constraints on both the
triplet vev $\langle \phi^0 \rangle$ and the quantities
$h_{ij}\langle \phi^0 \rangle$ in an analysis related to collider signals.

In this paper we point out the usefulness of looking for doubly-charged
scalars  in an $e^- e^-$ collider, in the radiative production channel. 
A linear collider with, say,
$\sqrt{s}~=~1$ TeV can operate for part of its run-time in the
electron mode, where certain signals can be remarkably free from backgrounds.
Also, this mode is perhaps ideal for exploring scenarios with
$\Delta L = 2$ couplings. As for doubly-charged scalars, their resonant
production in $e^- e^-$ as well as $\mu^- \mu^-$ have been already
studied \cite{Gunion, Raidal}.
However, resonant production of the $\f$ requires one to know
its mass with reasonable accuracy to start with, and tune the center-of-mass
energy of the colliding electrons accordingly. In addition, precise 
identification of a doubly-charged resonance will also depend on its
decay products \cite{GunWudka}, which depend on the parameters of the 
L-violating sector. In general, one can have the decays
\begin{itemize} 
\item $\f \longrightarrow W^{-}W^{-}$ 
\item $\f \longrightarrow l^{-}l^{-}$ 
\item $\f \longrightarrow W^{-}\phi^{-}$
\item $\f \longrightarrow \phi^{-}\phi^{-}$
\end{itemize}

Of these, the third mode, if kinematically possibly, is dominant as it
is driven by gauge coupling. However, a degeneracy among the triplet 
components is often a consequence of theories, albeit in a model-dependent
fashion. If we thus neglect the last two channels listed above, we still have 
the $W^{-}W^{-}$ and $l^{-}l^{-}$ channels, of which the first is controlled
by the triplet vev $v'$ and the second, by the coupling $h_{ll}$. When
the first mode is dominant, it requires careful analysis of the W-decay
products in order to isolate signatures of resonant production.
Furthermore, the analysis becomes complicated in case the $\f$ can
decay into a $\phi^{-}$ or a pair of them.

It is thus desirable to have supplementary channels in mind while looking
for doubly-charged scalars. With this in view, we have calculated the
rates for the process

$$
e^{-} e^{-} \longrightarrow \f \gamma \longrightarrow X \gamma
$$

\noindent
concentrating on the hard single photon in the final state. This photon
will be monochromatic if a doubly-charged resonance is produced, irrespective
of what it decays into. Furthermore, one is no more required to tune the
electron-electron center-of-mass energy at a fixed value \cite{skrai}. And 
finally, one can use such a study of monochromatic single photons in 
$e^{-}e^{-}$ collision to include the search for  doubly-charged
objects other than the scalar discussed above, an example being
bilepton resonances \cite{bileptons}.

For our analysis, taking the radiative production of the scalar $\f$ 
as the benchmark process, we concentrate only on the flavor diagonal 
coupling $h_{ee}$. 
Others, especially the non-diagonal couplings, are subject to
very stringent bounds from rare decay
processes of $\m^\pm$ and $\tau^\pm$ leptons. The rare decay studies
\cite{raredks}
provide bounds on the product of couplings as:
\bea 
h_{e\m} h_{ee} ~ < 3.2 \t 10^{-11} ~{\rm GeV}^{-2} M_{\h}^2 \\
h_{e\m} h_{\m\m}~< 2.0 \t 10^{-10} ~{\rm GeV}^{-2} M_{\h}^2 
\eea
The relevant bounds in our case are the following upper bounds, which
come from Bhabha scattering \cite{pheno}:
\bea 
h_{ee}^2 \sim 6.0 \t 10^{-6} ~{\rm GeV}^{-2} M_{\h}^2
\eea
and from $(g-2)_\m$ measurements \cite{g-2}:
\bea
 h_{\m\m}^2 \sim 2.5 \t 10^{-5} ~{\rm GeV}^{-2} M_{\h}^2
\eea
Much stringent upper bounds exist from the measurements of
muonium-antimuonium transition \cite{muonium} in the form of the 
product of the couplings:
\bea
h_{ee} h_{\m\m} \sim 2.0 \t 10^{-7} ~{\rm GeV}^{-2} M_{\h}^2
\eea
The above bounds allow for small value of doubly-charged scalar mass
with small coupling constant. In our numerical estimate, we have chosen the 
coupling strength to be $h_{ee}=0.1$ which is consistent with the limits 
obtained from Bhabha scattering \cite{LEPEXP,Bhabha} and also respects the 
most stringent bounds coming from muonium-antimuonium conversion 
results which for flavor diagonal coupling is 
$h < 0.44~M_\h~{\rm TeV}^{-1}$ at $90\%$ C.L. The latter bound however, 
can be relaxed in different new physics models \cite{NPM}.

As has been mentioned already, on-shell radiative production 
of a doubly-charged scalar gives an almost monochromatic 
photon of energy 
\bea
E_\g = \frac{s - M_{\f}^2}{2\sqrt{s}}
\label{egamma}
\eea
which stands out against the continuum background of the standard model
(SM). In the
discussions to follow based on our work, we show how, by simply tagging
on the isolated photon in the final state without bothering about the
the other associated particles, and looking at the line spectrum
superposed on the continuum background, leads to clear signals for the
production of a doubly-charged scalar.

Before we present the results of our analysis, it may be noted that
there exist  studies on the doubly-charged scalar in the context of
hadron colliders \cite{HatLHC} and 
in different modes of operation of linear colliders other than
$e^- e^-$ collision \cite{Raidal,ee} and $e^+ e^-$ mode \cite{godfrey}, 
such as the $e\g$ and $\g\g$ modes \cite{ephoton}. 
Although these works point out interesting ways of producing doubly-charged
scalars either singly or in pair, the latter are somewhat  
restrictive in mass reach, considering
the fact that a back-scattered photon carries a fraction
of the initial electron beam energy. Moreover, there is an unavoidable
kinematic suppression in pair-production in the $\gamma\gamma$ channel,
in spite of the spectacular enhancement of rates due to
the electric charge(s). Also, another possible channel for associated 
production of a doubly charged Higgs in $e^- e^-$ collisions has been
looked at in ref~\cite{bargeretal}.
In this work we show that we can probe the full
energy reach of the collider in its fundamental mode of running, and is
only restricted by the phase-space restrictions due to
the kinematic cuts used for the selection of events. Hardness and
transversality cuts on the photon in our final state are presumed to
avoid any confusion with initial state radiation (ISR) or beamstrahlung 
photons.

\begin{figure}[htb]
\begin{center}
\includegraphics[height=1.2in,width=4.5in]{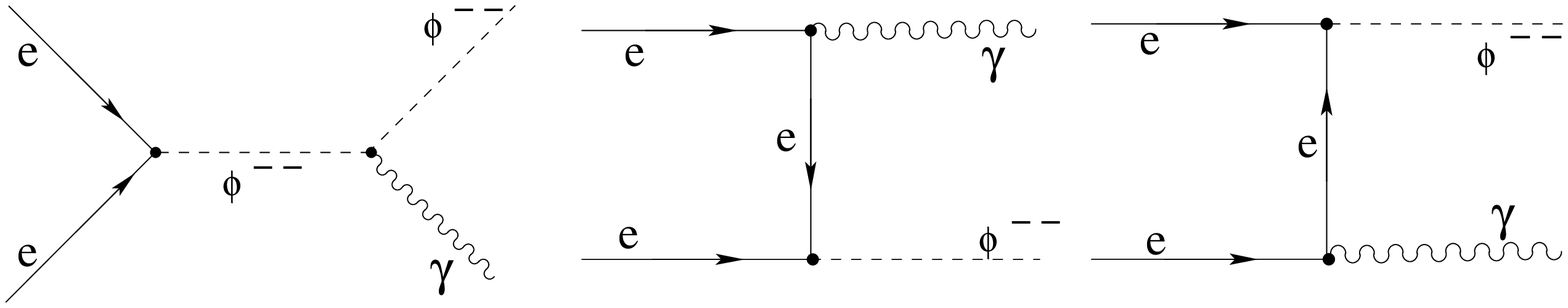}
\caption{\sl\small  Feynman graphs corresponding to single production of
 a doubly-charged scalar along with an associated photon in $e^-e^-$ 
collisions.}
\label{fgraphs}
\end{center}
\end{figure}

The Feynman diagrams contributing to the associated photon process
are shown in Figure \ref{fgraphs}. 
It is worth mentioning here that since they involve
clashing fermion lines, one has to be careful while writing down the
Feynman amplitudes and carefully handle the charge conjugation operator
appearing in the Feynman rules for the $eeH$ vertices. Using the
properties of the charge conjugation operator $C$, it is a matter of
simple algebra to write down the matrix amplitude squares for the
different graphs in Figure \ref{fgraphs}. The above process is also a clear
indicator of a $\Delta L=2$ process. We look at this process in the
context of a $\sqrt{s}= 1$ TeV linear $e^- e^-$ collider. As discussed
earlier, we consider decay of the doubly-charged scalar to like signed
leptons only and that too of the same flavor. We assume that the
$h_{ii}$ couplings are of equal strength and so the branching ratio
BR$(\phi^{--}\to l_i^- l_i^-)= 1/3$ for $i=1,2,3$.
In this work we have consistently chosen the coupling strength to be
$h_{ee} = 0.1$ and we calculate the decay width of the doubly-charged scalar 
assuming the decays to leptons only. In other words, we assume the triplet 
$vev$ to be very small, so it hardly contributes to the decay mode of
$\f \to W^- W^-$ which is directly proportional to the triplet
$vev$ squared ($v'^2$) and we have also neglected the other possible decay 
modes of $\f$ as pointed out earlier. We find that the total decay width 
obtained is very miniscule ($\sim 1.2$ GeV) for a 1 TeV scalar mass
when compared to the machine energy. So we can use the narrow-width 
approximation and consider on-shell production of the doubly-charged scalar 
and its subsequent decay to $e^- e^-$.

The major SM background that contributes to the above process is the
radiative Moller scattering process: 
$$e^- + e^- \to \gamma + e^- + e^-$$ 
which, although a continuum background, could {\it prima facie} be
large enough to wash away the monochromatic peak. The event selection
criteria, therefore, are largely aimed at suppressing this continuum
background.

We impose the following set of cuts. Since we
do not want the final state particles to be too close to the beam pipe,
one needs to have a rapidity cut on the final state particles:
$$|\eta(e^-)| < 3.0 ~~~~~~~~{\rm and}~~~~~~~~~|\eta(\gamma)| < 2.5$$
We also demand a hard photon in the final state, which is the main
focus of our work. This is obtained by imposing a cut on the minimum 
photon energy:
$$E(\gamma) > 20 ~{\rm GeV}$$
A minimum energy is also demanded for both the final state electrons:
$$ E(e^-) > 5 ~{\rm GeV}$$
The above criteria help us in suppressing the continuum background to
a considerable extent.
\begin{figure}[htb]
\begin{center}
\includegraphics[width=2.5in]{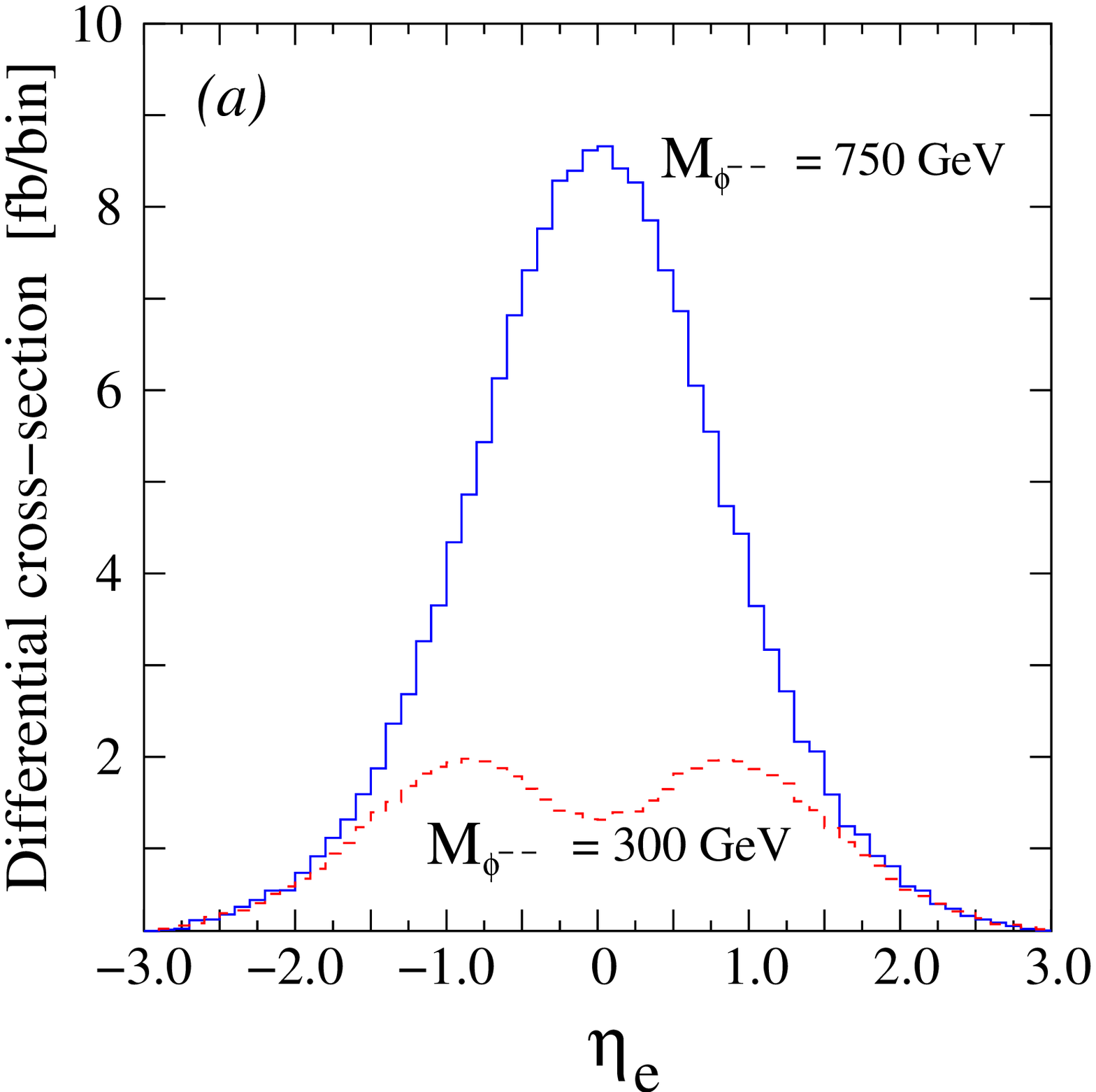}
\includegraphics[width=2.6in]{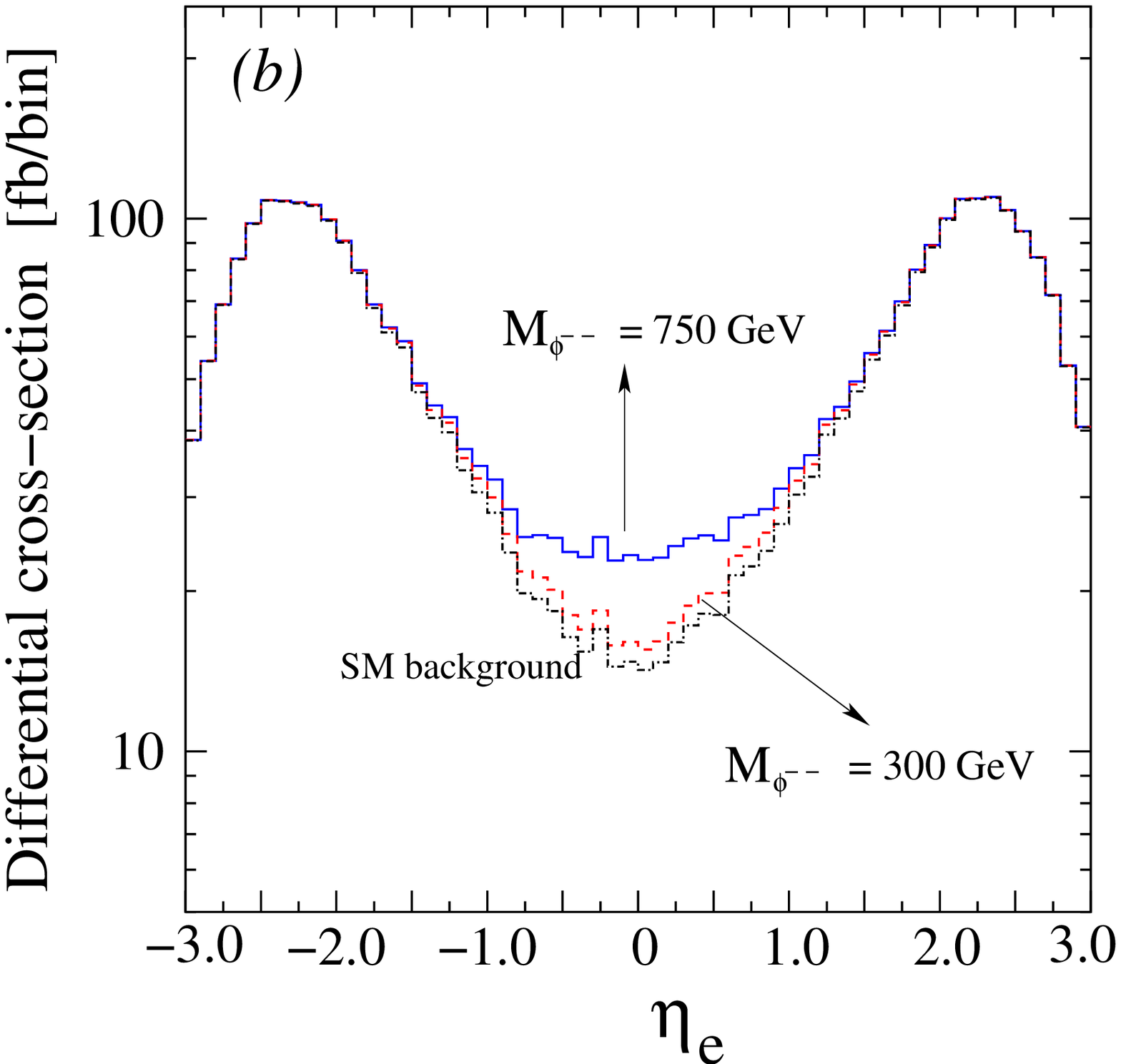}
\includegraphics[width=2.6in]{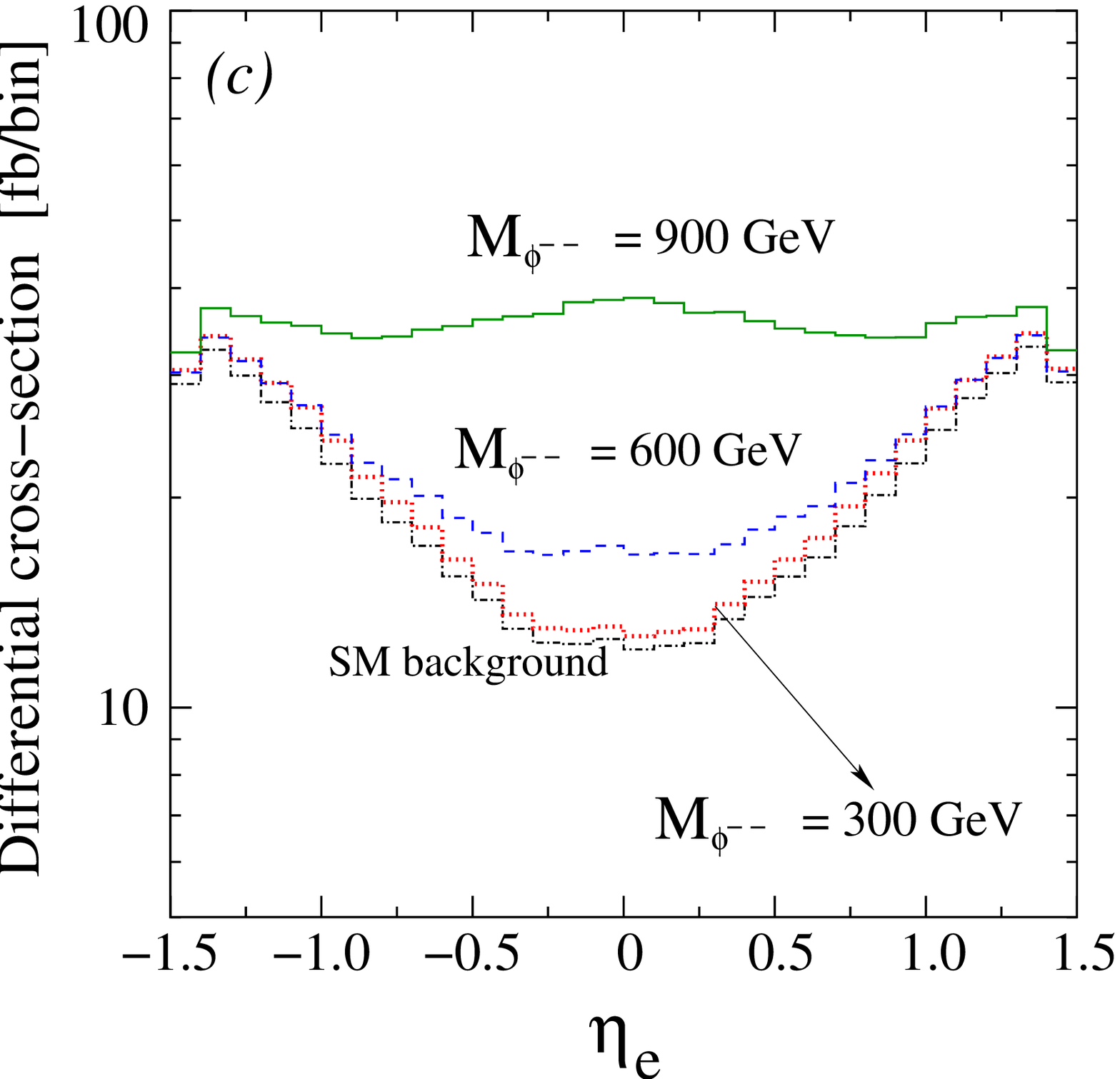}
\caption{\sl\small Differential cross-sections for (a) signal with
$|\eta_e| < 3.0$, (b) {\it signal+background} and SM background with
$|\eta_e| < 3.0$ and (c) {\it signal+background} and SM background with
$|\eta_e|< 1.5$, against electron rapidity $\eta_e$.}
\label{rapidity}
\end{center}
\end{figure}
Finally we would like the detectors to register and resolve events for 
the different particles and hence all the final state particles should
be well separated in space and satisfy:
$$\delta R > 0.2$$
where $(\delta R)^2 \equiv (\Delta \phi)^2 + (\Delta \eta)^2$ with
$\Delta \eta$ and $\Delta \phi$ respectively denoting the separation in
rapidity and azimuthal angle for the pair of particles under
consideration. Using the above cuts we make an estimate of the SM
background and the signal. The background has been generated using the
MadEvent Monte Carlo generator \cite{madE}. 
It is found that the background is quite
large compared to the signal with a cross-section of about $3216~fb$.
In Figure \ref{rapidity}. we show the distribution for the differential
cross-section as a function of the electron rapidity. It is worth 
noticing that the
SM background is symmetric in the binwise distribution for $\eta(e^-)$
of the final state electron and is peaked away from the central region
of the rapidity distribution (Fig \ref{rapidity}(b)), 
which means that they are more in the forward direction. This is 
expected due to the strong t-channel radiative contribution to 
Moller scattering. On the same curve we also show the distribution for 
the {\sl signal+background} for two different values of the doubly-charged 
scalar mass, $m_{\f}= 300$ GeV and $750$ GeV. In contrast to the
background, we find that the signal is peaked at the central region, 
which is clear
from Fig \ref{rapidity}(a) where we have plotted the {\sl signal} alone 
and this peaking
\begin{figure}[htb]
\begin{center}
\includegraphics[height=3.1in,width=3.1in]{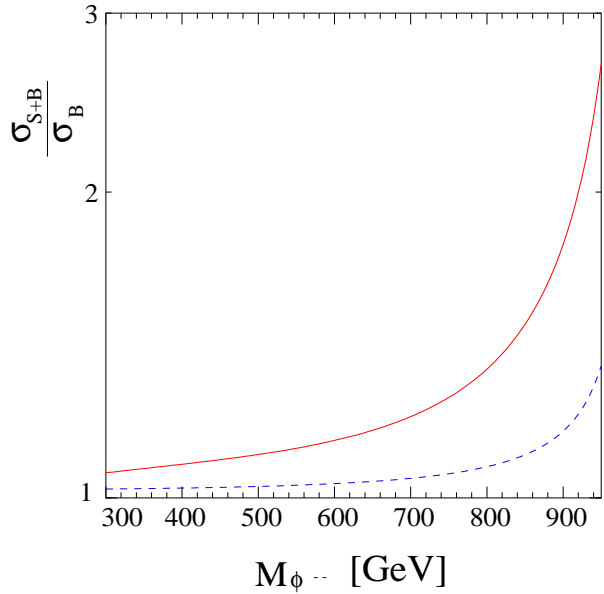}
\caption{\sl\small Illustrating the ratio of signal+background with the SM
background. The broken line (blue) corresponds to the cut $|\eta_e|<3.0$
and the solid line (red) corresponds to the cut $|\eta_e|<1.5$}
\label{ratio}
\end{center}
\end{figure}
becomes more pronounced as the mass of the $\f$ increases.
Infact for the low mass there does not seem to be much difference in the
rapidity distribution but as the mass is increased to a higher value,
the difference shows up with the signal peaked centrally. This can be
easily understood if we look at the kinematics of production. The higher
the mass of the $\f$, the less boost it will have and hence the
decay products will come out back-to-back. Using this as a cue, we can
actually implement a more stronger cut on the rapidity of the electron
which will throw the signal into prominence. We find that a cut of
$$|\eta(e^-)| < 1.5$$
is good enough to kill the background by about $80\%$ while the signal
goes down only by about $45\%$ for low masses ($\sim 300$ GeV) and by 
about $10\%$ for large masses ($\sim 900$ GeV) for the doubly-charged
scalar.  With the kind of luminosity expected at future linear colliders
we expect that the event rates would be considerable even if low mass
state is realized and although the signal might drop by $40\%-50\%$,  
it will still stand out against the much reduced background due to 
this cut. To highlight this we implement the cut and plot the 
\begin{table}[htb]
$$
\begin{array}{|c|c|c|c|c|}  \hline
{\rm Cut~on~}\eta_e&\multicolumn{2}{c|}{|\eta|<3.0} &\multicolumn{2}{c|}{|\eta|<1.5}\\\hline
M_{\h}(GeV)& \s_B (fb)& \s_S (fb) & \s_B (fb) &\s_S (fb)\\\hline
  300 &       &  65.5 &        &   35.4 \\
  400 &       &  72.7 &        &   47.9  \\
  500 &       &  84.8 &        &   62.5  \\
  600 & 3216.0& 105.9 &  614.0 &   84.1  \\
  700 &       & 146.1 &        &  122.2  \\
  800 &       & 235.6 &        &  204.7  \\
  900 &       & 524.9 &        &  468.8  \\\hline
\end{array}$$
\caption{\sl\small Cross-sections for signal and the SM background 
corresponding to the two choices of rapidity cut on the final state 
electron.}
\end{table}

distribution in Fig \ref{rapidity}(c). It is clearly visible how the 
signal is 
enhanced compared to the background with this cut. We have demonstrated 
the {\sl signal+background} for three different masses for the 
doubly-charged scalar, $m_{\f}= 300,600,900$ GeV. We also show the 
significance of the cut in Fig \ref{ratio} by plotting the ratio of the 
cross-sections of the {\it signal+background} ($\s_{S+B}$) with 
the {\it SM background} ($\s_B$), for both
the choices of the cut on electron rapidity, as a function of the 
doubly-charged scalar mass. The graph clearly highlights the
enhancement in the signal to background ratio and reflects the increase
in the cross-section as we go higher in scalar mass. In Table 1, 
we list the cross-sections (rounded off to the nearest integer) 
for different masses corresponding to the respective cuts imposed on 
the electron rapidity for a more quantitative outlook. 
\begin{figure}[htb]
\begin{center}
\includegraphics[height=2.65in,width=2.6in]{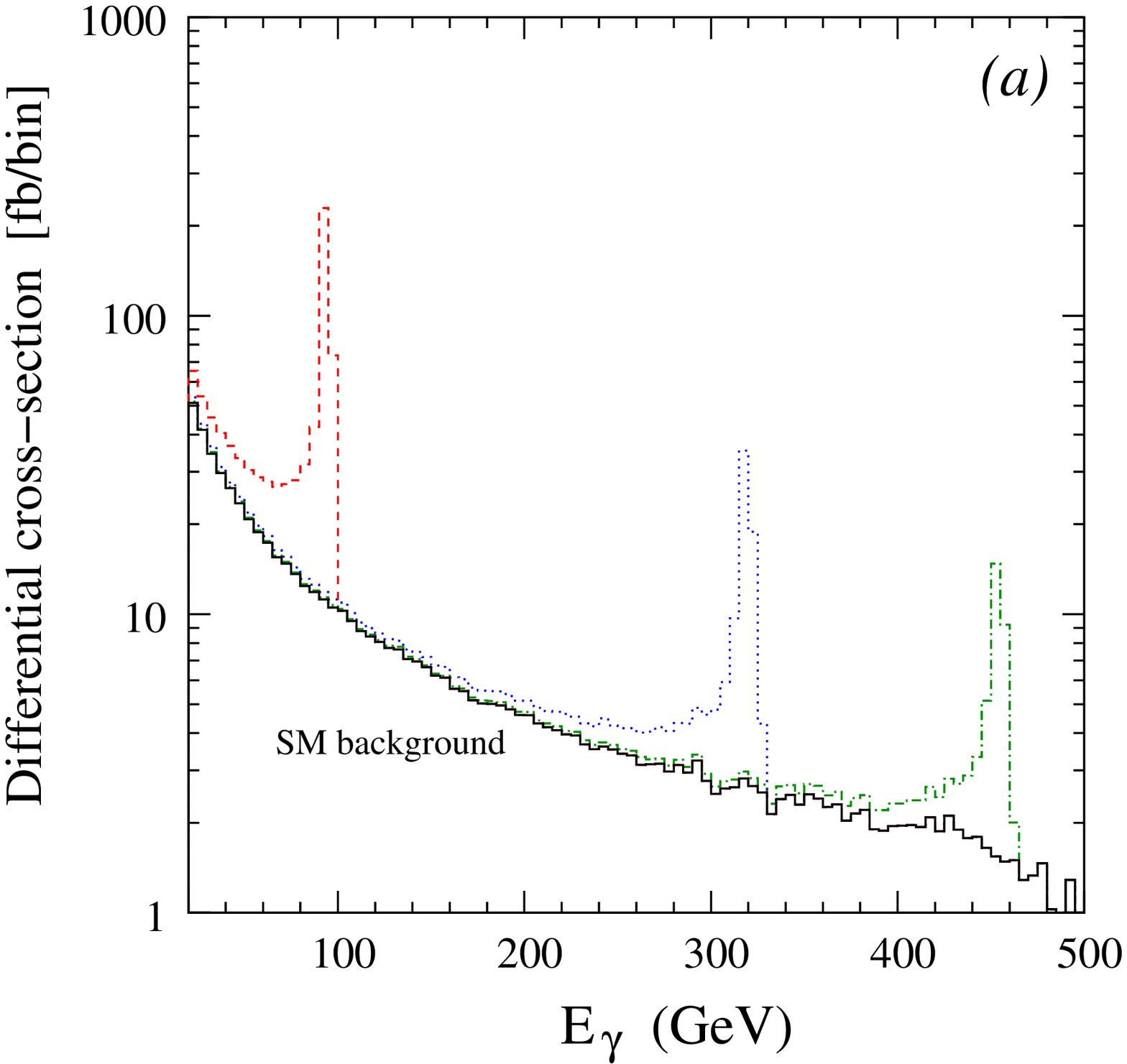}
\includegraphics[height=2.6in,width=2.6in]{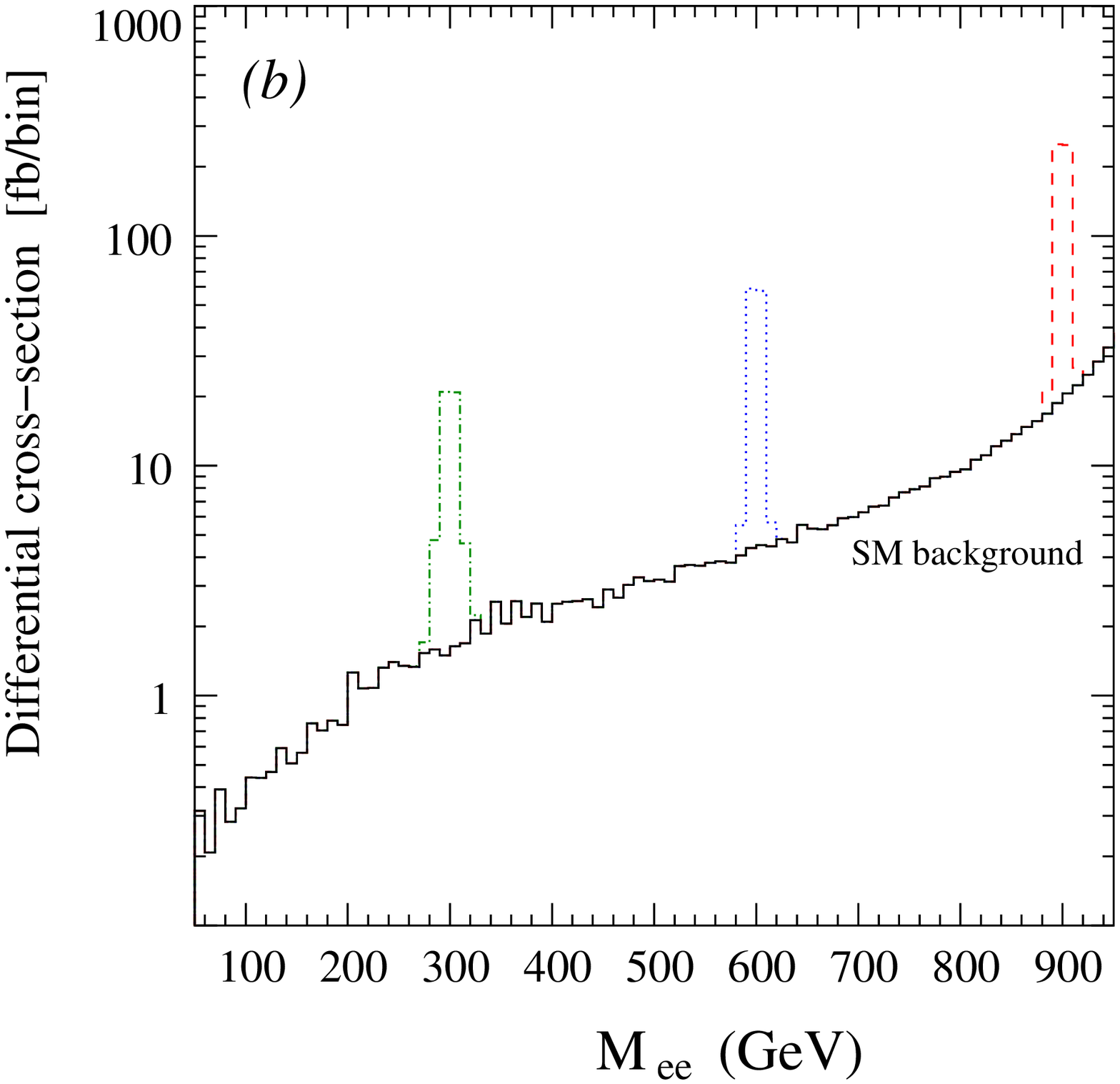}
\caption{\sl\small Differential cross-sections against 
(a) photon energy $E_\g$ and (b) invariant mass of electron pair $M_{ee}$. 
The dash-dot-dash (green) line corresponds to $M_{\phi^{--}} = 300$ GeV, 
dotted (blue) line corresponds to $M_{\phi^{--}} = 600$ GeV and the 
dashed (red) line corresponds to $M_{\phi^{--}} = 900$ GeV
respectively. The binsize is chosen to be 5 GeV in (a) and 10 GeV in (b).}
\label{resonance1}
\end{center}
\end{figure}

Next, we focus on the main trigger, {\it viz.} the photon. In Fig
\ref{resonance1}(a) we show the distribution of the photon energy, 
where we have
superposed the differential cross-section for {\it signal+background} 
in each bin over the SM background. A pronounced peak can be seen in
the photon energy distribution, due to the monochromaticity of
the photon, corresponding to the recoil energy
against the scalar resonance through the relation of Eq.\ref{egamma}.
To make our analysis realistic, we have smeared the photon energy by a
Gaussian function whose half-width is guided by the resolution of the
electromagnetic calorimeter \cite{smear1,smear2}:
$$ \frac{\Delta E}{E} = \frac{14\%}{\sqrt{E}}$$
Moreover, we have fully incorporated the effects of ISR \cite{isr} which 
often results in substantial broadening of the peak, due to the spread 
in the effective center of mass energy available for our process. We
have used CompHEP \cite{chep} to include ISR effects for the SM
background.
We show the resulting peak for three choices of scalar mass (300, 600, 900
GeV). Alternatively, in Fig \ref{resonance1}(b), we also show the invariant 
mass distribution of the $ee$ pair for the above choice of parameters 
and as expected the distribution peaks corresponding to the mass of 
scalar. Here the half-width of the Gaussian function used for smear is
\cite{smear1,smear2}:
$$ \frac{\Delta E}{E} = \frac{15\%}{\sqrt{E}} + 0.01 $$ 

\begin{figure}[htb]
\begin{center}
\includegraphics[height=3.1in,width=3.1in]{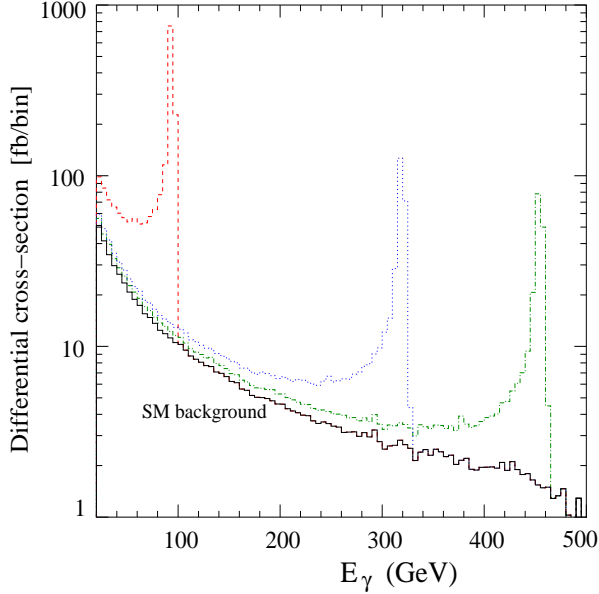}
\caption{\sl\small Differential cross-sections against photon energy $E_\g$
 when $\phi^{--} \to X$(anything). The dash-dot-dash
(green) line corresponds to $M_{\phi^{--}} = 300$ GeV, dotted (blue) line
corresponds to $M_{\phi^{--}} = 600$ GeV and the dashed (red) line 
corresponds to $M_{\phi^{--}} = 900$ GeV respectively. The binsize is 5
GeV.}
\label{resonance2}
\end{center}
\end{figure}

In Fig \ref{resonance2} we plot the energy distribution of the 
photon once again. But
here we assume that we do not have prior knowledge of the decay products
of the doubly-charged scalar. In other words, we only look at the final
state hard transverse photon in          
$$e^- e^- \to \g + \f \to \g + X$$
The distribution again shows peaks corresponding to the
recoil against the massive scalars, irrespective of the knowledge of the
decay products of the scalar. In fact our signal here receives a
relative boost as it 
is not suppressed by considering any further decay since the 
$BR(\phi^{--}\to X)= 100\%$. Through Fig \ref{resonance1}. and 
Fig \ref{resonance2}. we have shown that a
single associated photon will show peaks in its energy distribution over
the continuum background of SM, if a doubly-charged scalar is produced
with mass $M_{\f} < \sqrt{s}$. We only demand that no electron in the
final state can have a rapidity whose absolute value is greater than
1.5.  The fact that looking at a single 
photon against the backdrop of a continuum background makes it possible to
identify a LFV ($\Delta L=2$) process in a model independent way, makes
this signal worth studying at a future $e^- e^-$ collider and running
the linear collider in this mode. 

Since the rates for the signal depend directly on the $eeH$ coupling 
squared, we can make an estimate of the strength of the coupling which 
can give substantial rates for identification of the peaks in the 
\begin{figure}[htb]
\begin{center}
\includegraphics[height=3.5in,width=3.5in]{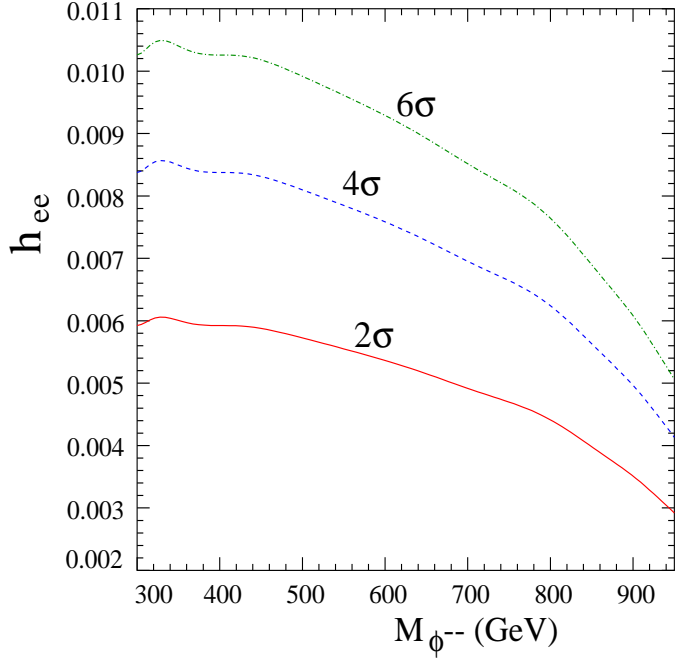}
\caption{\sl\small Illustrating the reach of the coupling constant at 
which the resonances in the $E_\g$ distribution can be identified over the
fluctuations in the SM background. The assumed luminosity is 
100 fb$^{-1}$.}
\label{coupling}
\end{center}
\end{figure}
photon energy distribution. We do this for the case when $\f\to X$ and
in Fig \ref{coupling}. We show the strength of the coupling for which 
the peaks would 
stand out against the fluctuations in the SM background at $2\s, 4\s$ 
and $6\s$ level. In our analysis we have assumed a luminosity of 
$\mathcal{L} = 100~{\rm fb}^{-1}$, easily achievable at future linear
colliders. We have also done this analysis based on the smear in photon
energies due to finite resolution of the detector as well as ISR effects, 
as mentioned earlier. A bin of width 10 GeV about the signal peak in photon
energy has been identified for each $M_{\f}$. We
then look at the fluctuations in the SM background in that bin and compare 
the corresponding rate for the signal. This procedure has been repeated 
for different coupling strengths. The fact that we are not looking at
any specific final state arising from $\f$ decay improves the reach of
this search channel. Our analysis suggests that the peaks in the photon
distributions will be distinguishable for coupling strengths as low as
$0.006(0.010)$ for scalar mass of 300 GeV at $2\s (6\s)$ level and to
about $0.0034(0.006)$ for scalar mass of 900 GeV at $2\s (6\s)$. This
estimate far overwhelms the simple method of comparing total rates of
signal and background and restricting the coupling strength in the
parameter space. It is also worth mentioning that if one is to assign a
certain order of detection efficiency $\epsilon$, with the final states
then the above reach atmost scales by $\epsilon^{-1/4}$. 
However, if a direct resonance is excited then that
would invariably translate into a much stronger probe of the coupling
strength \cite{Gunion}. Nonetheless, our analysis is not dependent on
the tuning of the $\sqrt{s}$ of the machine to hit a resonance 
and hence serves as a more robust proposition. For luminosity higher
than what we have used, this reach can be further enhanced.

To summarise, the cleanliness of central photon detection at a high energy
linear collider can be very helpful in identifying a doubly-charged
scalar. While it is true that final states such as $(\mu^-\mu^-)$ can
be completely background-free, the peaks in the hard photon energy can
be helpful in two ways. First, one does not need to tune the two
electron beams, and can therefore work without a prior knowledge of the
$\f$ mass. Secondly, this method is shown to work even if the $\f$
dominantly decays into states that are not clean enough for the
resonance to be identified. Thus, as soon as one succeeds in reducing
the SM backgrounds (using, for example, the electron rapidity cut), one
can clearly see $\Delta L=2$ interactions, just by looking at the
accompanying hard photon. Not only doubly-charged scalars but also more
exotic resonances such as bileptons are amenable to detection in
this manner.

\vskip 5pt
{\bf Acknowledgments:}
SKR would like to thank Anindya Datta and R. Srikanth for helpful 
comments. We also acknowledge illuminating discussions with
Aseshkrishna Datta.

\end{document}